\documentclass[twocolumn]{emulateapj}
\usepackage{graphicx}
\usepackage{wasysym}
\usepackage{graphicx}

\slugcomment{Submitted to ApJ}

\shorttitle{Two-Dimensional Map of an Exoplanet}
\shortauthors{Majeau et al.}

\begin{document}

\title{A Two-Dimensional Infrared Map of the Extrasolar Planet HD~189733b}

\author{Carl Majeau\altaffilmark{1,2}, Eric Agol\altaffilmark{1}
and Nicolas B. Cowan\altaffilmark{1,3,4}}
\altaffiltext{1}{Department of Astronomy, Box 351580, University of Washington,
    Seattle, WA 98195}
\altaffiltext{2}{Columbia University, New York, NY}
\altaffiltext{3}{Northwestern University, 2131 Tech Drive, Evanston, IL 60208}
\altaffiltext{4}{CIERA Postdoctoral Fellow}

\email{n-cowan@northwestern.edu}

\begin{abstract}
We derive the first secondary eclipse map of an exoplanet, HD 189733b,
based on Spitzer IRAC 8 micron data.  We develop two complementary techniques for 
deriving the two dimensional planet intensity:  regularized slice mapping and 
spherical harmonic mapping.  Both techniques give similar derived 
intensity maps for the infrared day-side flux of the planet,
while the spherical harmonic method can be extended to include phase variation data
which better constrain the map.
The longitudinal offset of the day-side hot spot is consistent with that found 
in prior studies, strengthening the claim of super-rotating winds, and eliminating 
the possibility of phase variations being caused by stellar variability.  
The \emph{latitude} of the hot-spot is within 10.1$^{\circ}$ (68\% confidence) of 
the planet's equator, confirming the predictions of general circulation models 
for hot Jupiters and indicative of a small planet obliquity.
\end{abstract}


\section{Introduction}

One of the great challenges in studying extrasolar planets is that we cannot
directly resolve the surfaces of these bodies.  This is a common problem in
astronomy, and one solution is to use occultations or eclipses to spatially
resolve astronomical sources.  Eclipse mapping has been 
applied variously to map stars \citep[e.g.][]{1993Sci...262..215S,
1998A&A...332..541L}, accretion disks \citep[e.g.][]{1985MNRAS.213..129H}, 
planetary satellites \citep[e.g.][]{1975AJ.....80...56A}, 
dwarf planets \citep[e.g.][]{1999AJ....117.1063Y,
2001AJ....121..552Y}, and unresolved
radio sources \citep[e.g.][]{1966ApJ...146..646T, 1967ApJ...150..421T}.  Here we 
present the first eclipse map of an extrasolar planet.

The secondary eclipse is the decrease in flux that occurs when a planet passes behind 
its host star.  The eclipse ingress and egress contain information about the spatial 
distribution of the planet's day side flux. 
For example, a planet with a brighter western hemisphere has an apparent delay
in the times of ingress and egress, and a modified shape of the ingress and egress.  
This delay in eclipse timing due to a zonally asymmetrical planet was predicted by 
\cite{Williams_2006} and first observed by \cite{Knutson_2007} and \cite{Agol_2010}.
In addition to this artificial time offset, the ingress/egress morphology may be used
to map exoplanets as pointed out in \cite{Rauscher_2007}.  

Longitudinal variations in the brightness of the planet cause a variation in
the brightness of the planet with orbital phase.  This orbital modulation 
has been observed in transiting \citep{Knutson_2007, Knutson_2009a, 
Knutson_2009b} and non-transiting systems \citep{Cowan_2007, Crossfield_2010}.  
\cite{Cowan_2008} described how to invert such phase variations into a low-resolution
{\it longitudinal} map of a planet; the first such map was presented in \cite{Knutson_2007}.

\subsection{The Hot Jupiter HD~189733b}
The short-period transiting jovian planet HD 189733b \citep{Bouchy_2005}
is one of the first planets to have been detected in infrared emission
during secondary eclipse \citep{Deming_2006}, and has been
studied in depth since that detection \citep{Knutson_2007,Grillmair_2008,
Charbonneau_2008, Knutson_2009a, Agol_2010,Desert_2011}.

We apply our mapping techniques to the seven secondary eclipse measurements
of \cite{Agol_2010} and, in the case of spherical harmonic mapping, the phase 
measurement presented in \cite{Knutson_2007}.  These data have been corrected 
for detector effects and stellar variability, and the seven eclipse 
light curves were binned in phase to 644 points.  To this we appended the 
phase variation data from \cite{Knutson_2007}, binned by 275.

\section{Eclipse mapping} \label{eclipsemap}
\subsection{Model Assumptions}
Three conditions need to be satisfied in order to construct an eclipse map:
the planet emission pattern must be static, the time of superior conjunction must be known, and
the planet limb-darkening must be negligible.

A static diurnal temperature pattern is required for any form of thermal mapping. 
Most general circulation models (GCMs) for hot Jupiters indicate $<1$\% variability 
\citep{Showman_2002, Cooper_2005, Showman_2009, DobbsDixon_2010}, while observations of 
multiple secondary eclipses of HD~189733b indicate that the eclipse depth does not 
vary significantly, $<2.7$\% over a period of more than 500 days \citep{Agol_2010}.  
Hence the static map assumption is likely justified for this planet.

Eclipse mapping can be applied provided that the time of superior conjunction is known 
to a few seconds, which is beyond the precision of radial velocity measuremenents.  
We \emph{assume} a circular orbit, hence pegging the time of eclipse 
to the time of transit which is precisely known.  This assumption can then be checked by 
comparing the phase-function and secondary eclipse maps, and seems to apply for HD~189733b 
\citep{Agol_2010}.

Ignoring the effect of limb darkening is required so that the
surface brightness of a location on the planet does not change as the planet
rotates. This assumption is not required for our slice-mapping technique, since in 
that case we implicitly neglect changes in the planet's phase.  
\emph{Regardless, }limb darkening is thought to be very weak for hot Jupiters, about 20\% linear 
limb-darkening, especially in the wavelength of the observations (8 micron),
so this assumption is likely reasonable for HD 189733b, as we show below.

In addition to these assumptions about the planet, we assume some things of the data.
Any stellar variability must be corrected 
for, which may not be possible if it has a period close to that of the planet.  We have
corrected for stellar variability using the correlation between optical and infrared
flux variations as discussed in \citet{Agol_2010}.  Detector 
ramp and other detector-related effects are corrected for as described in
\citet{Agol_2010}.  Additionally, all light 
curves henceforth are taken to have the stellar flux subtracted off,
so that measurements when the planet is completely occulted have zero value.  

We describe two approaches to planet mapping:
(1) slice mapping (\S \ref{slice_section}); (2) spherical harmonic mapping 
(\S \ref{ylm_section}).  

\begin{figure*}[htb]
\includegraphics[width=\hsize]{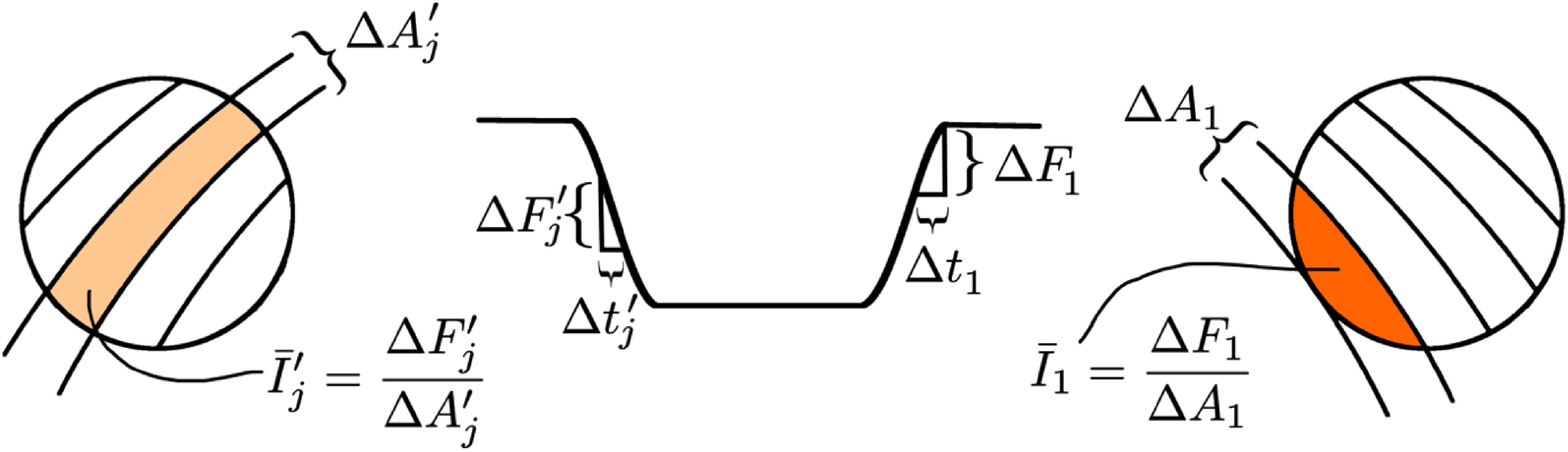}
\includegraphics[width=\hsize]{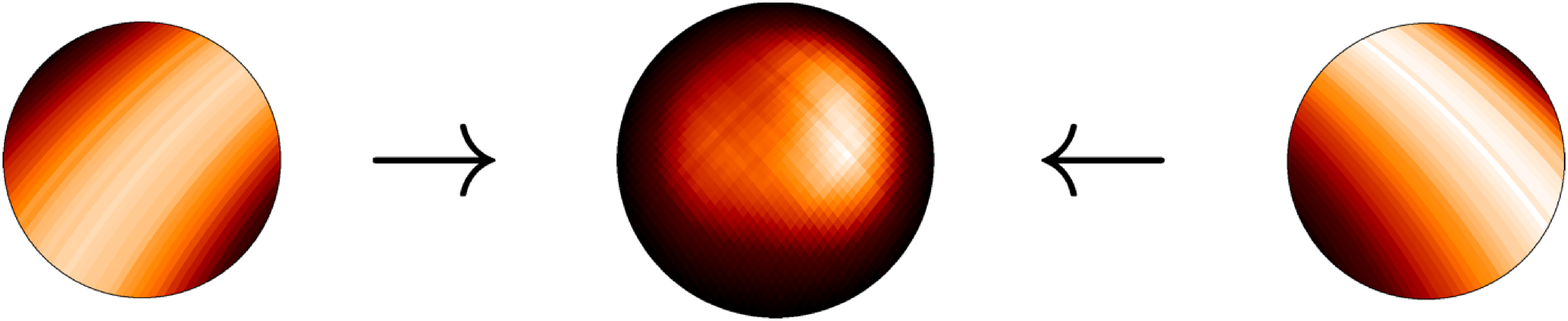}
\caption{Top: diagram demonstrating the slice mapping technique. If one adopts the convention that North points in the direction of the planet's rotational angular momentum, then western regions of the planet are occulted first during ingress, and revealed first during egress (for a tidally-locked planet the rotational and orbital angular momenta are identical). If North points up and East to the right, then ingress is on the right and egress on the left; time therefore runs from right to left. Bottom: Ingress and egress maps (right and left), as well as a combined map (center) of HD~189733b at 8~$\mu$m. The image is centered on the sub-observer point and the black regions are 50\% as bright as the white.} \label{slice_map}
\end{figure*}

\subsection{The Slice Method} \label{slice_section}
During a time interval within ingress/egress, we can compute the area of the planet 
that disappears/appears behind the star if we assume a circular orbit.  The fraction of 
flux that disappears during that time interval is then proportional to the mean surface 
brightness of the area that disappeared. This is the basis of our first mapping technique.

As a planet passes behind its star, the light curve flux, $F(t)$, at any given time may be
associated with the roughly crescent-shaped portion of its day side currently visible to the earth.
It is thus proportional to the planet's light distribution function, $I(\theta,\phi)$, 
integrated over the visible area, $A(t)$.
If we take two successive observations at times $t_{j-1}$ and $t_j$, the change in observed flux, 
$\Delta F_j = F(t_j) - F(t_{j-1})$, is due solely to the change in area 
$\Delta A_j = A(t_j) - A(t_{j-1})$ times the intensity averaged over that area.
The average intensity, $\bar I_j$, over the slice with area $\Delta A_j$ is then given by 
$\bar I_j=\Delta F_j/\Delta A_j$.  

We partition the planet's ingress and
egress into observations at a finite number of intervals, $\Delta t_j$, and from these calculate each 
$\bar I_j$, yielding two `slice maps': 1D maps of the planet day side. 
This process and its results for HD 189733b are shown in Figure 
\ref{slice_map}.

Once these maps were found, we combined them into a single 2-dimensional map of the 
planet, forming a grid of the ingress and egress slices.  This step of the inversion requires
a prior since if the ingress and egress maps are each made from $N$ slices, 
the combined map contains roughly $N^2$ cells.  We assume that the planet's intensity 
distribution varies smoothly, which is imposed via linear regularization, 
in which a solution is chosen that is consistent with the data and
has the smallest differences between adjacent slices.  We use a goodness of fit parameter 
of $\chi^{2} + \lambda\sum_{i} ({\bar I_{i}}-{\bar I_{i-1}})^{2}$.  We chose a
regularization coefficient, $\lambda = 100$,  to give a reduced chi-square of order
unity, resulting in the map of HD 189733b shown in Figure \ref{slice_map}.

\subsection{Spherical harmonic mapping} \label{ylm_section}
Our second method for extracting a map from the eclipse light curve uses spherical harmonics. 
We applied this to two data sets: 1) only the secondary eclipse data; 2) both the secondary 
eclipse and phase-function data.  

\begin{figure}
\includegraphics[width=\hsize]{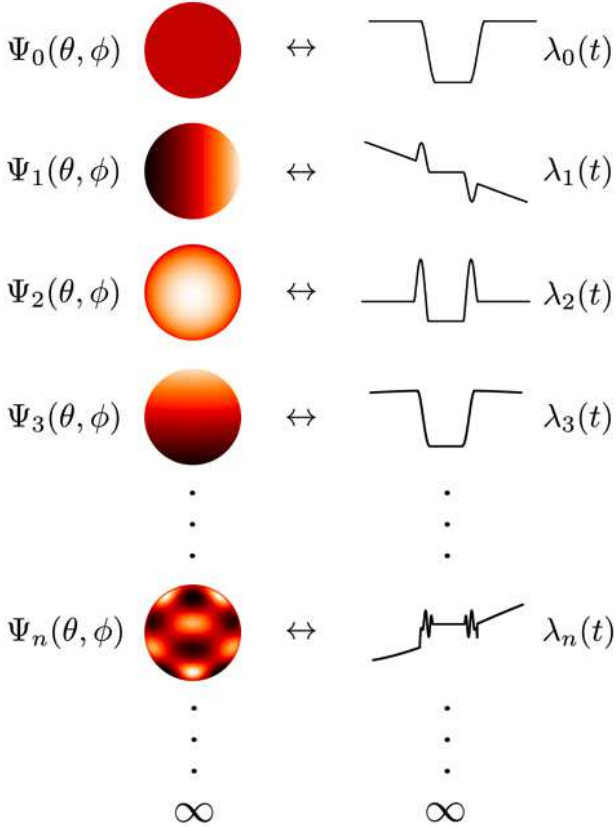}
\caption{ A few spherical harmonics and their corresponding `harmonic light curves,' 
$(\lambda_0, \lambda_1, \lambda_2,...)$.}
\label{spherical_harmonic_technique}
\end{figure}

Taking $I(\theta,\phi)$ to be the intensity map of a transiting extrasolar planet, we may approximate it 
to an arbitrary degree of accuracy by truncating its harmonic sum at some finite number of 
terms, $0 \le l \le l_{max}$, with $-l \le m \le l$.
We compute the light curve that would be observed if the planet's map consisted solely of 
each harmonic, and then solve for the coefficients of each harmonic light curve from the data via linear 
regression.  We implemented the light curve calculation using the {\it HEALPix} IDL package,
which contains routines for the evaluation and projection of spherical harmonics
\citep{Gorski_2005}.
The map is then formed by transforming back to the harmonics and summing the result. 
This transformation is portrayed graphically in Figure \ref{spherical_harmonic_technique}, 
and the results of its application to observations of HD 189733b can been seen in Figure 
\ref{hd189733_map}.  This is essentially a two dimensional generalization of the Fourier 
decomposition technique for phase mapping developed in \citet{Cowan_2008};  note that
the 2D image shown in \cite{Knutson_2007} only contained longitudinal information, while
the latitudinal depdence was arbitrarily chosen.

We generate the light curve for the real spherical harmonics, $\Psi_i$ (where
$i$ labels the harmonics used in our inversion) by painting a
planet with a particular harmonic and then stepping through the planet's orbit for 
each of the observed orbital phases.  At each phase the harmonic is integrated over the
visible portion of the planet to get the total flux that would be
observed.  We integrate over the visible area by finding a function with curl
equal to the harmonic and integrate it around the area's boundary,
which is simply the concatenation of two known arcs, using Green's
Theorem.  This technique yields both the planet phase variation and secondary
eclipse shape for each harmonic, as shown in Figure \ref{spherical_harmonic_technique}.

With a set of $n$ spherical harmonic light curves, $\lambda_1(t),...,\lambda_n(t)$,
we decompose the observed light curve, $F(t)$, by computing the
the coefficients $c_1,...,c_n$ as $c_i = \int F(t)\lambda_i(t)dt.$ 
These translate directly to the map's coefficients for the harmonics
$\Psi_1,...,\Psi_n$. By adding the terms together we form a best-fit map of the
entire planet, the closest approximation we can reach using the
chosen harmonics, shown in Figure~\ref{hd189733_map}.

For the HD 189733b Channel 4 IRAC map, we chose to use only the first four harmonics 
$(l, m) = {(0, 0), (1, -1), (1, 0), (1,1)}$ which gave a reduced chi-square
of the fit close to unity.  Thus, the maps only contain
a dipole variation in flux, which corresponds to the ``hot spot" feature
seen in simulations and inferred from the phase function.

\begin{figure}
\includegraphics[width=\hsize]{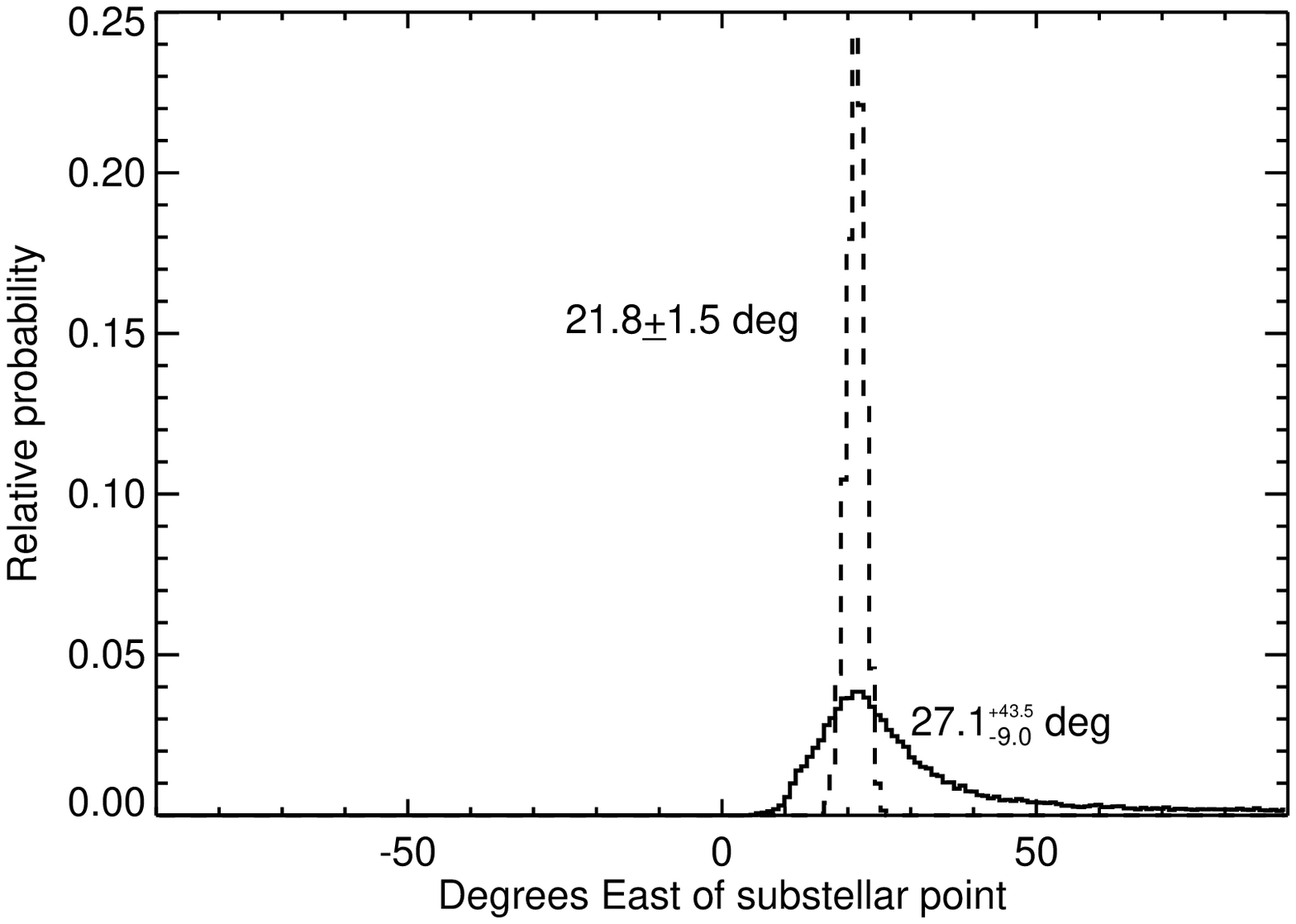}
\includegraphics[width=\hsize]{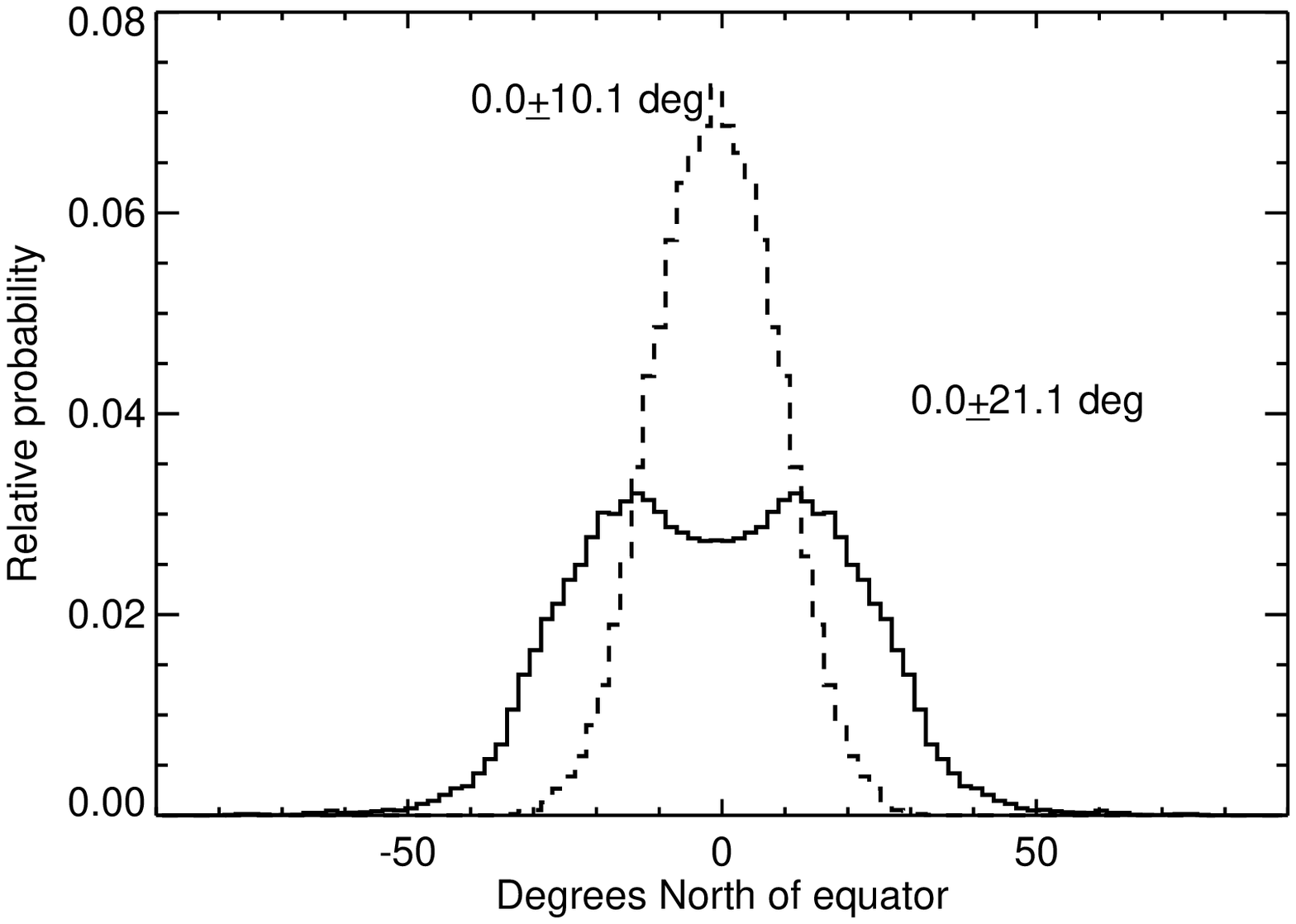}
\caption{Probability distribution for the longitude and latitude of the hot spot in the $l_{\rm max} = 1$ spherical harmonic maps based on the secondary eclipse (solid lines) and secondary eclipse plus phase function (dashed lines). Angles are measured in the rotation frame of the planet, with the origin at the sub-stellar point.  It is not known whether the orbital angular momentum of the system is tilted towards or away from the observer, so we are only able to constrain the distance of the hot spot from the equator, not specify whether it is in the northern or southern hemisphere.}
\label{hot_spot_location}
\end{figure}

To quantify the statistical uncertainty on the measured hot spot
location, we ran a Markov chain using the spherical harmonic fits.  { Beginning with a 
uniformly bright planet in a 4-dimensional vector space corresponding to the first four 
harmonic coefficients, we attempted jumps with magnitude based upon the uncertainty found in 
the initial decomposition for each coefficient, and a jump factor adjusted every 100 
steps to achieve and acceptance rate of $\approx 44$\%, with the acceptance
criterion based on the Metropolis-Hastings algorithm.
The longitudinal and latitudinal positions of the hottest point on the planet model 
was recorded for each of the 25,000 iterations, and the resulting distribution is plotted 
in Figure \ref{hot_spot_location}.}

It is not known whether the orbital angular momentum of the system is tilted towards or away from the observer, so we are only able to constrain the distance of the hot spot from the equator, not specify whether it is in the northern or southern hemisphere. Accounting for this degeneracy, we find a hot spot latitude within $21.1$ degrees of the equator for the secondary eclipse inversion, albeit with a slight preference for off-equatorial locations, and within $10.1$ degrees of the equator for the eclipse-plus-phase inversion.

The corresponding
longitude was $27.1^{+43.5}_{-9.0}$ deg East and $21.8\pm 1.5$ deg East for the
eclipse or eclipse plus phase inversion, respectively.   Consequently,
the phase function gives significant leverage in constraining
both the longitude and latitude of the hot spot.  

Thermal phase variations are completely ``blind'' to meridional (N--S) brightness variations, so it is worth explaining how these data improve the constraints on the latitude of the hot spot. Due the planet's non-zero impact parameter, the star's limb scans diagonally across the planet, so eclipse mapping is unable to break the degeneracy between zonal (E--W) and meridional concentration of the dayside flux. The ingress and egress maps (Figure~\ref{slice_map}) scan in complementary directions, however, and therefore could distinguish between a zonal and meridional \emph{offset} of the hotspot.  Nevertheless, there remains a partial degeneracy between zonal/meridional concentration and hot spot offset.  By measuring the zonal brightness concentration and longitude of the hot spot, phase variations break this degeneracy, leading to stronger constraints on the latitude of the hotspot.

In any case, we
estimate that both hot spot longitude and latitude are subject to $\approx 15$\% systematic
errors due to neglect of $l>1$ modes in the inversion based
on our simulated data, as described next.

\section{Application to Simulated Data}

We created simulated light curves using a two dimensional map at constant
pressure from a GCM for a hot Jupiter planet 
\citep{Showman_2008}, to which we applied the spherical harmonic mapping 
technique.  We took a snap shot of the temperature at 100 mbar from a 
simulation, converted it to surface brightness
in the Rayleigh-Jeans limit, and then created a simulated light curve
from the images.  The light curve was computed for identical orbital
and size parameters as HD 189733a,b and for the same range of phase
that was observed (about 1/3 of the planet's orbit prior to secondary
eclipse), and we assumed that the planet
has a fixed temperature pattern with no limb-darkening, as well as
a circular orbit and spherical bodies.  We carried out six different 
versions of the simulated light curves:  1) we first
decomposed the temperature map into spherical harmonics up to order
$l_{max}$, and then used the map based on only these harmonics to
compute a light curve with no noise; 2) we computed the same light
curve, but with noise added at the level of the observed data from \cite{Agol_2010}; 
{ 3) same as 2, but we shifted the hot spot North in latitude by
12, 23, 40 and 60 degrees;} 4) we
computed a light curve using the full temperature map; 5) we added noise to the
light curve using the full map; 6) we carried out an identical test
as version 1, but with 20\% linear limb-darkening.   We found that for $l_{max}=1$,
we could recover the correct coefficients of the spherical harmonics
in version 1, which shows that the code performed as expected.
When noise was added, we found a scatter in the derived parameters
at the 10\% level.  With $l_{max}=2$, we found that the derived
spherical harmonics became very poorly constrained, likely due to
degeneracies in the harmonic light curves, so for the remainder
of the analysis we used $l_{max}=1$ when extracting the map.  
{ For version 3, we ran 100 simulations at each latitudinal offset,
and found that the mean of the recovered latitudes matched the input
value, with a scatter varying from 12 deg RMS at the equator to 6 deg
nearer to the pole.} For version 4, we
found that the amplitude of the spherical harmonics relative to
the $(0,0)$ component had systematic errors as large as $\approx 50$\%.
However, the amplitude of the spherical harmonics relative to the 
{\it day-side flux} inferred from the map agreed to better than 15\%,
while the location of the hot spot differed by
$\approx 10-15$\% in longitude and latitude.    
For version
5 we found that the statistical error did not change much for
$l_{max}=1$, still at the 10\% level.  For version 5, we find that
20\% limb darkening affects the derived coefficients by less than 5\%.

Based on these tests, we are confident in our inference of the 
longitude and latitude of the planet hot spot the dipole amplitude
relative to the day-side flux at the $\approx 15$\% level.  { Additionally, we confirm the validity of our assumption to ignoring the effects of limb darkening for hot Jupiters in 8 micron.}

\section{Discussion and Conclusions} \label{discussion}
\subsection{Slice vs. Spherical Harmonic Eclipse Maps}
\begin{figure}
\includegraphics[width=\hsize]{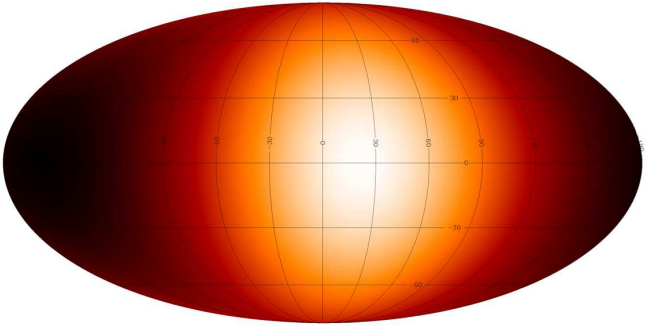}
\caption{A Molleweide projection of the 2D infrared map of HD 189733b with 
$l_{max}=1$; latitude and longitude lines are spaced every 30$^\circ$. 
Coordinates are centered on the sub-stellar point and intensity values are relative
(white is brightest, black is 30.2\% as bright).}
\label{hd189733_map}
\end{figure}

The fact that the slice map (Figure \ref{slice_map}) and
spherical harmonic map (Figure \ref{hd189733_map}) give qualitatively
similar results is reassuring as the two techniques have
very different assumptions and potential systematic errors.
Both maps favor a hot spot that is offset slightly latitudinally;
however, when the phase function is included in the analysis, this
constrains the spot to be located more closely to the equator.
This is due to the fact that the phase function provides a
strong constraint on the longitudinal variation of
the planet brightness, and thus does not allow as much freedom
in varying both the longitude and latitude to fit the secondary
eclipse.

The slice map has the disadvantage that it does not account
for the rotation of the planet or the tilt of the planet, nor
can it be used in conjunction with the phase variation of the
planet, { resulting in the north-south asymmetry visible in the slice map above absent from the harmonic mapping results.}  In addition, it is difficult to derive uncertainties
on the intensity, which are highly correlated due to the regularization.

The spherical harmonic method has the advantage of being able
to easily account for these effects, but the disadvantage that it assumes
to a large degree that the planet intensity varies smoothly as we
are constrained to use small $l_{max}$, in this case $l_{max}=1$.

\subsection{Phase versus Eclipse Mapping}
Phase mapping \citep{Cowan_2008} offers more leverage on the longitudinal 
brightness distribution of the planet than does eclipse mapping.  This 
is because the same change in flux is spread out over an orbital period, $P$,
rather than the duration of ingress/egress, $\tau_{ing}$, allowing for greater 
signal-to-noise constraints given the same photometric uncertainties.  
The ratio of these two timescales is $\tau_{\rm ing}/P \propto R_{p}/(\pi a)$
($\approx $0.4\% for HD 189733), 
where $R_{p}$ and $a$ are the planetary radius and orbital semi-major 
axis, respectively.  Whenever it is feasible to obtain photometry throughout 
an entire orbit (or a sizable fraction thereof), phase curve inversion 
will provide the best estimate of the hottest and coldest local stellar 
times.  This will be especially true as one tries to apply these techniques 
to smaller planets with longer periods.

Eclipse mapping offers three significant advantages over phase mapping, however: 

1) Eclipse mapping is sensitive to to the latitudinal intensity distribution on a planet
if it has non-zero impact parameter as it passes behind the disk of its star.

2) The planet's brightness map is assumed to be static over the eclipse duration, 
$\tau_{\rm ecl}$, which is much shorter than that assumed for phase mapping, $P$, and thus
influenced less by other forms of variability.  The ratio of these two timescales
is $\tau_{\rm ecl}/P \propto R_{*}/a$ ($\approx$ 3\% for HD 189733).  While there is 
no clear evidence yet for weather 
on transiting planets, it will certainly be the case that the intensity distributions
of eccentric planets as seen by an observer will vary throughout an orbit.
Eclipse mapping therefore offers the only prospect for constraining the intensity 
distributions of this intriguing class of objects. 

3) There is no theoretical limit to the spatial resolution one can achieve with 
eclipse mapping, although there may be some degeneracy as we have found.  This is 
in stark contrast to 
phase mapping, where one runs into fundamental degeneracies with odd harmonics after only 
5 terms \citep{Cowan_2008}.      

\subsection{Summary} 
We developed two techniques for converting an observed secondary eclipse light curve 
into a two-dimensional map of a planet.

The spherical harmonic formalism is a robust way to make an exoplanet map from secondary 
eclipse light curves, and can accommodate observations of thermal phase variations.  
Slice maps may also be constructed at ingress and egress, then combined via regularization.  
While this technique seems to produce comparable day-side maps for the data at hand, 
it is not as robust or versatile as the spherical harmonic technique.

We applied both mapping techniques to the transiting planet HD~189733b.  We find that 
the primary day-side hotspot is offset to the East by 21.8(1.5)$^{\circ}$, in agreement 
with the phase variation measurements of \cite{Knutson_2007} and the eclipse timing 
constraint from \cite{Agol_2010}. The
data indicate a hot spot within $10.1^\circ$ of the equator.  The fact that the hottest point lies close to the equator
is consistent with a small obliquity of the planet as oblique rotators are predicted 
to show strong latitudinal contrast \citep{Langton_2007};  this conclusion is
also consistent with theoretical computations which predict a small obliquity
for hot Jupiters \citep{Fabrycky_2007,Levrard_2007}.
Taken together, these studies strongly favor super-rotating winds in the 
vicinity of the 8 micron photosphere, as predicted in GCMs
\citep{Cooper_2005,Showman_2008,Showman_2009,Rauscher_2010,Burrows_2010,
Showman_2011}.  
The mapping codes used in this paper, written in IDL, are 
available from the authors upon request;  however, the assumptions that go into
this planet mapping technique need to be verified for other planets studied.

\acknowledgments
We thank Adam Showman for providing the temperature map used to
test our inversion technique. We thank Julien de Wit for pointing out the erroneous caption for Figure 1 in the original paper.
This work is based in part on observations made with the
Spitzer Space Telescope, which is operated by the Jet Propulsion
Laboratory (JPL), California Institute of Technology under contract
with NASA. Support for this work was provided by NASA
through an award issued by JPL/Caltech.
We acknowledge support from NSF CAREER grant AST-0645416.

\end{document}